**Running the COVID-19 marathon: the behavioral adaptations in mobility and facemask in the first 27 weeks of pandemic in Seoul, South Korea**

**Authors:** *Jungwoo Cho, PhD[1,2], Yuyol Shin[1,2], Seyun Kim[1], Namwoo Kim[1], Soohwan Oh[1], Haechan Cho[1] Yoonjin Yoon, PhD[1]\**

**Affiliations**

[1] Transportation Research and Urban Engineering Lab, Department of Civil and Environmental Engineering, KAIST, Daejeon, South Korea.

[2] These authors contributed equally to this work

\*Correspondence to: yoonjin@kaist.ac.kr

**Abstract**

Battle with COVID-19 turned out to be a marathon, not a sprint, and behavioral adjustments have been unavoidable to stay viable. In this paper, we employ a data-centric approach to investigate individual mobility adaptations and mask-wearing in Seoul, South Korea. We first identify six epidemic phases and two waves based on COVID-19 case count and its geospatial dispersion. The phase-specific linear models reveal the strong, self-driven mobility reductions in the first escalation and peak with a common focus on public transit use and less-essential weekend/afternoon trips. However, comparable reduction was not present in the second wave, as the shifted focus from mobility to mask-wearing was evident. Although no lockdowns and gentle nudge to wear mask seemed counter-intuitive, simple and persistent communication on personal safety has been effective and sustainable to induce cooperative behavioral adaptations. Our phase-specific analyses and interpretation highlight the importance of targeted response consistent with the fluctuating epidemic risk.

# 1. Introduction

Mobility intervention has been embraced as one of the most effective and immediate control measures since the early days of COVID-19 pandemic, and universal quarantines such as lockdown and shelter-in-place have shown measurable success. Sjödin et al.[1] conducted a modeling study to show that the lockdown duration could be shortened when household size and time spent outside one's home were brief. Since household size could not be easily changed, the study concluded that strict quarantine measures are the key to reduce its duration. Leung et al.[2] showed that reproduction number has dropped below 1 after lockdowns in 14 Chinese provinces, and the fatality risk could also be decreased. Based on the Susceptible-Infectious-Recovered model, the study projected that premature relaxation of lockdown measures would increase the cumulative case count exponentially. López and Rodó[3] used a stochastic Susceptible-Exposed-Infectious-Recovered to project the epidemic progression after lockdown, suggesting that another larger scale infection wave is likely if lockdown measures were alleviated prematurely and abruptly.

Although strong mobility stringency can minimize uncertainties arising from individual differences to achieve prompt and well-coordinated response, high social and economic costs have been inevitable. Suppressing personal mobility was found to deepen economic inequalities[4,5], generate disconnects among communities[6], and even cause mental illness[7]. As the pandemic prolonged, concerns over accumulating costs and exit strategies have deepened. Phased exit strategy, prioritizing people at the highest risk, and managing misinformation were suggested to reduce unintended cost in the post-lockdown period[8].

Another key control measure of massive testing and contact-tracing has shown promising outcomes in several countries including South Korea[9]. However, a recent study found that such approaches are feasible only when case numbers remain at a manageable level, and the success of prevention measures depends on the population's behavioral changes in response[10]. Hellewell et al.[11] stated that contact tracing and isolation without mobility reduction might not be sufficient to contain the outbreak under plausible scenarios.

Even if the testing-contact-tracing has been effective to contain the outbreak, it cannot serve as a direct incentive to induce behavioral adaptations. The main motivation of our study is consistent with the aforementioned studies, and we aim to evaluate the role of mobility when case numbers remain at a traceable level. Several recent studies also emphasized the need to understand behavioral aspects and perception changes regarding policy adherence[12], and the importance of mobility data analyses for a broader understanding of the efficacy of public communication and social distancing interventions[13]. Despite the urgent need for adaptive mobility policy targeting behavioral adaptations, there are limited academic and technical resources available. It is encouraging, however, that an increasing number of researches utilize large-scale near-real time datasets to study the interrelationship among mobility change, social impact, and policy response[5,6,14].

In this paper, we employ data-centric approaches to trace the evolution of mobility behavioral adaptation and its relationship to epidemic progression in Seoul, South Korea for the first 189 days (27 weeks). As the 4th country infected with COVID-19, the nation had been the worst-affected country outside China in the beginning[15]. However, the nation has managed to control the infection rate under 20.99 per million population since January 20, 2020 without lockdown-type mandates. The nation's capital Seoul also has been able to contain the maximum daily confirmed cases under 52 or 5.34 per million population in the same period with minimal mobility restrictions.

Due to the time-dependent and transient nature of epidemic progression and mobility patterns, modeling their relationship over the entire 27 weeks is meaningless. As the first step, we propose a framework to identify epidemic phases based on the case count as well as the geospatial spread of COVID-19. The mobility patterns are represented with the reductions in subway ridership (public transit) and traffic volume (private mode) compared to the same period of 2019, given the mobility network under full operations. Association between mobility and epidemic progression is then modeled as linear models in six epidemic phases using a changing point detection technique. Seasonality specific to mobility including the day-of-week and the time-of-day effect is also evaluated to discover common focus in mobility decisions. Lastly, findings from the statistical analyses are assessed with risk perception and facemask use in a collective manner.

## 2. Method

### 2.1. Determination of epidemic phases

The major transitions in epidemic progression were determined by detecting the major structural changes in daily case counts with a changing point detection (CPD) method of breakpoint detection algorithms[16,17]. The algorithms simultaneously detect the major interventions in sequential data, and have been widely used in multiple disciplines including finance[18-21], genetics[22,23], climate change[24,25], social science[26-28], ecology[29], and geology[30,31]. In our analysis, the intercept-only linear model of $y = \beta_0 + \epsilon$ was assumed to detect the day of major structural change, and used in the COVID-19 case count analysis and the mobility seasonality analysis.

Although the breakpoint analysis method detects the structural changes in a robust manner, sequential data in general represent long-term trends in a retrospective manner, and is sensitive to the timespan under investigation. To complement such shortcomings, we propose to measure the geospatial progression of COVID-19 in addition to the confirmed case count. Conceptually, it is intuitive that quantifying spatial dispersion provides useful insights for an infectious disease like COVID-19, as the risk cannot be the same when 100 new cases are located in a single location versus spread evenly across a region. Methodologically, we employed two topological measures of grouped distance ($d_g$) and Hausdorff distance ($d_H$). Since both measures represent the distance between two sets, one can reduce the two-day changes into single metrics. If the daily case sequence represents the long-term trend, the topological measures are designed to capture the short-term momentum.

Given the set of locations exposed by COVID-19 patients, which we call the *contact locations*, let $P_t = \{p_t : contact\ locations\ on\ day\ t\}$ be the set of all contact locations of day *t*. The grouped distance between $P_t$ and $P_{t+1}$ is defined as $d_g(P_t, P_{t+1}) = \frac{1}{|P_t| \cdot |P_{t+1}|} \sum_{p_t \in P_t, p_{t+1} \in P_{t+1}} \|p_t - p_{t+1}\|$. It is the average distance to reach from a point in $P_t$ to points in $P_{t+1}$, representing proximity of two sets. While small $d_g$ indicates $P_t$ and $P_{t+1}$ are more concentrated in a local cluster, large $d_g$ indicates they are more scattered. Hausdorff distance is defined as $d_H(P_{t+1}, P_t) = \max_{p_{t+1} \in P_{t+1}} \min_{p_t \in P_t} \|p_{t+1} - p_t\|$, which is the maximum distance to reach from $P_{t+1}$ to $P_t$. Large $d_H$ is observed when at least one new case occurred far from the existing mass.

Both measures are bounded between 0 and the length of the major axis of the geographical boundary. Theoretically, both measures can reach zero if the existing mass was a single cluster, and new cases also belong to the same cluster. Likewise, both measures can reach near the maximum when new cases are located in the opposite corner from the existing mass. Once the contact locations start to spread and fill up the region, $d_g$ will remain close to half of the maximum, while $d_H$ will be much smaller than that. More formally, $d_H > d_g$ means that there exists at least one location on day $t+1$ that contains none of the contact locations of day $t$ within the radius of $d_g$. Conversely, $d_g > d_H$ means that all locations of day $t+1$ contain at least one of the locations of day $t$ within $d_g$. In our context, new cases need to be closer to the existing mass when existing cases are scattered with less concentration, yielding $d_g \gg d_H$, which is the *geospatial peak*. In the analysis, the metric $d_g - d_H$ was used to identify the geospatial peak period compared to the geospatial contraction and expansion.

Once the epidemic phases of trigger, escalation, peak, and de-escalation phases were established using both the count interventions and the topological metric, the relationship between mobility and epidemic progression was modeled as linear models by individual phases. In the bivariate linear model, the response variable is the mobility reduction, and the COVID-19 count is the regressor.

**2.2. Data**

There are four main datasets used in the study – the hourly subway ridership and traffic volume, COVID-19 case counts and contact locations, risk perception of COVID-19 infection, and universal mask adoption rate. While the mobility and COVID-19 data were direct observations, risk perception and mask adoption rate were based on Likert scale responses sourced from a series of cross-sectional surveys. Subway ridership included the number of passengers getting on and off at 275 stations of subway lines 1 through 8, aggregated in hourly intervals. Traffic volume included the number of vehicles at 100 sensor stations located in the city's major road links. To incorporate the strong seasonality in mobility demand, changes were measured in the amount of reduction compared to the same period of 2019 as $1 - \frac{value\ of\ 2020}{value\ of\ 2019}$, by matching the closest day-of-week. For instance, Friday, February 7, 2020, was matched with Friday, February 8, 2019. COVID-19 daily confirmed case counts and contact tracing information were sourced from multiple sources including Kaggle COVID-19 open data, KCDC (Korea Centers for Disease Control and Prevention), and Seoul Metropolitan government. Daily data values are 7-day simple moving averages unless noted otherwise.

Regarding risk perception and universal mask adoption rate, we resorted to several independent survey results due to the lack of direct observation data and its highly qualitative nature. Two types of risk perception—overall risk of COVID-19 and risk of personal infection— were obtained from biweekly surveys conducted by Hankook Research. Overall risk of COVID-19 represents the percentage of people who responded that the spread of the pandemic within the nation was "very" or "somewhat" serious. Risk of personal infection represents the percentage of people who responded that the likelihood of contracting the virus was more than or equal to 50%[32].

The rate of universal mask adoption was based on five independent survey outcomes[33-37], which captured the proportion of the surveyed having worn masks whenever necessary. Additional details and availability of data used are included in the supplementary.

## 3. Results

### 3.1. Determination of COVID-19 phases

In Figure 1-a, the breakpoint analysis results of COVID-19 case count trend are shown with four interventions, yielding five phases divided on day 50, 82, 128, 156. In Figure 1-b, $d_g - d_H$ is shown with four transition days on 29, 82, 106, 169. When combined, we have 7 transition points in total. Although not detected in case counts, day 28 can be considered as the transition from a trigger to an escalation phase. Day 50 is the start of the peak phase in the first wave. Day 82 is the transition from the peak to a de-escalation. Day 106 divides the third phase into the de-escalation of the first wave and the escalation of the second wave. Day 128 is the start of the second wave peak. Although one can speculate day 156 is the start of a de-escalation, the case counts are not low enough compared to the first de-escalation. Moreover, $d_g - d_H$ stayed positive until day 168, suggesting the second peak had continued closer to 168. Although the validity of spatial dispersion results after day 169 is limited due to the contact location availability (see Supplementary Note 1), it is unlikely that the city had entered a de-escalation phase with $d_g - d_H$ being mostly positive. As results, we defined six epidemic phases as the trigger (1-29), escalation-1 (30-50), peak-1 (51-82), de-escalation (83-106), escalation-2 (107-128), and peak-2 (129-189).

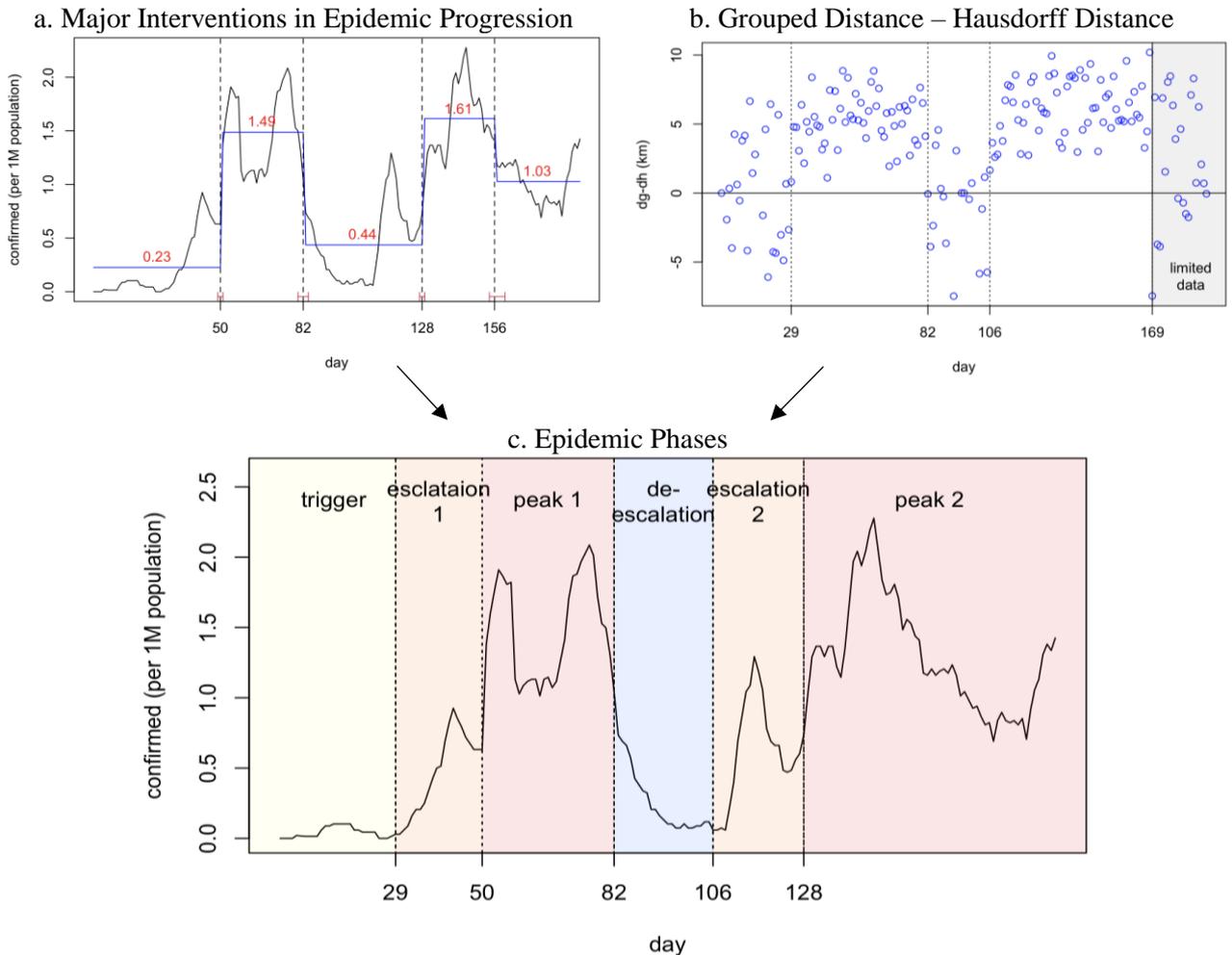

**Figure 1. Epidemic phase determination. a.** breakpoint analysis results of major structural changing points yielding five phases; **b.** periods of geospatial peak vs geospatial expansion/contraction yielding five phases; **c.** resulting six phases.

## 3.2. Association between epidemic progression and mobility

Linear model results are shown in Figure 2. In subway ridership, an addition of a single case resulted in an additional 4.1% reduction in escalation-1, which converted to a 0.5% recovery in escalation-2. In traffic volume, the additional reduction was 1.3% in escalation-1, and a 0.1% recovery in escalation-2. Between two peaks, the constant 40.8% subway ridership reduction in peak-1 was decreased to 20.9% in peak-2. Traffic volume shows recovery from an 8.8% reduction to 4.2% with a 0.1% additional recovery per additional new case.

The results provide several useful insights. The reduction ratio between public transit (subway)

to private trip mode (traffic volume) was approximately 4 to 1, manifesting the strong shift in mode preference. One can also infer that the reductions of approximately 40% in public transit use and 10% in traffic volume are possible upper-bounds of voluntary mobility adjustment, and equivalent to 32 day delay to enter the de-escalation phase. Strong positive linearity in escalation-1 combined with the maximum reduction throughout peak-1 indicates the public's willingness to embrace behavioral adjustment with rising infection risk, even without stringent mobility policy. However, conversion from the strong positive in the first escalation to marginally negative linearity in the second indicates significant changes in the public's perception of mobility as the main means of personal safety.

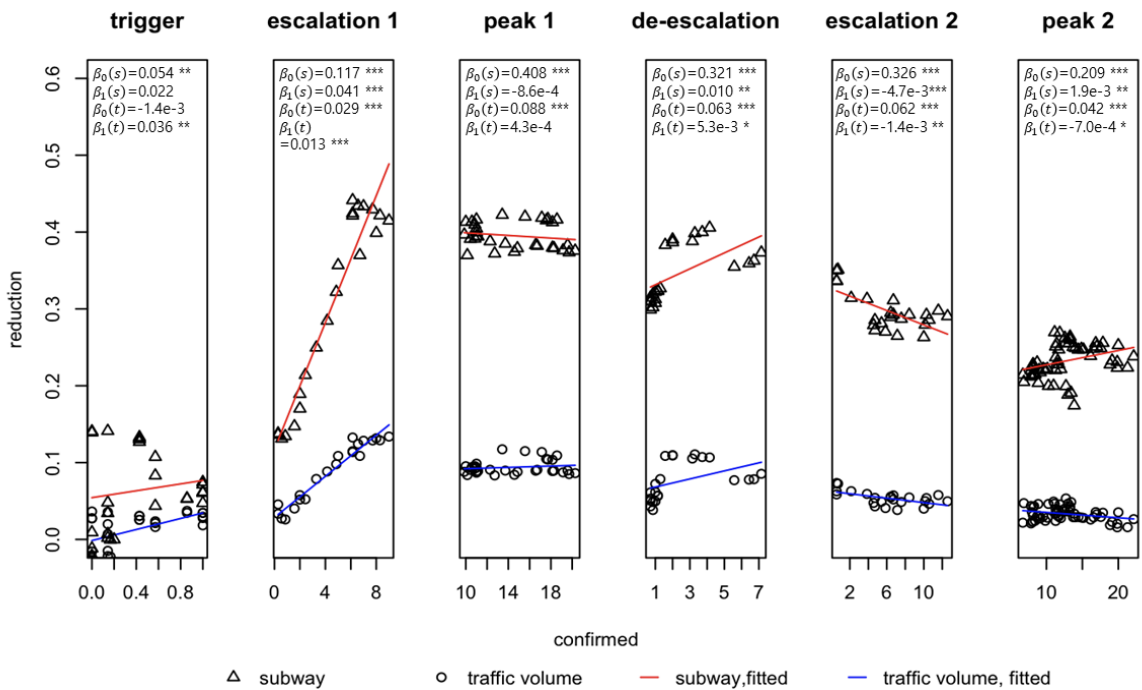

**Figure 2. Linear model fit of *mobility reduction vs. confirmed cases* of six epidemic phases.** $\beta_0(s)$ and $\beta_1(s)$ are the intercept and coefficient of *subway reduction vs. confirmed cases* model. $\beta_0(t)$ and $\beta_1(t)$ are the intercept and coefficient of *traffic reduction vs. confirmed cases* model. ($\beta$ values with *, **, *** indicate that the p-values are smaller than significance levels of 0.05, 0.01, 0.001, respectively.)

### 3.3. Seasonality in mobility

Since mobility patterns are embedded with strong seasonality, mobility patterns were evaluated for its time-of-day and day-of-week effects. As shown in Figure 3, the overall trends of two trip modes are similar except in magnitude, and public transit demand continues to recover as early as day 63 (week 9). The subway ridership shows reductions at least three times larger than traffic volume, which agrees with our findings in section 3.2. It is also worth noting that the subway has continued operations during the epidemic, serving more than 50% of the usual demand. Despite the strong preference shift to private mode, public transit has remained a vital urban mobility option, especially to those with limited mode choices.

It is notable that the first major interventions of two modes agree on day 34 (week 5), which is before any major government response is put in place. Moreover, weekends, nighttime, and afternoon trips show larger reductions compared to weekdays and commute hours. Under low policy stringency, such outcomes strongly suggest that early mobility reduction had been driven by proactive individual adaptations with a common focus on reducing non-essential trips while preserving the essential commute trips in private mode. Continuous recovery after the de-escalation signifies changes in public perception regarding the role of mobility, which agrees with our findings in section 3.2.

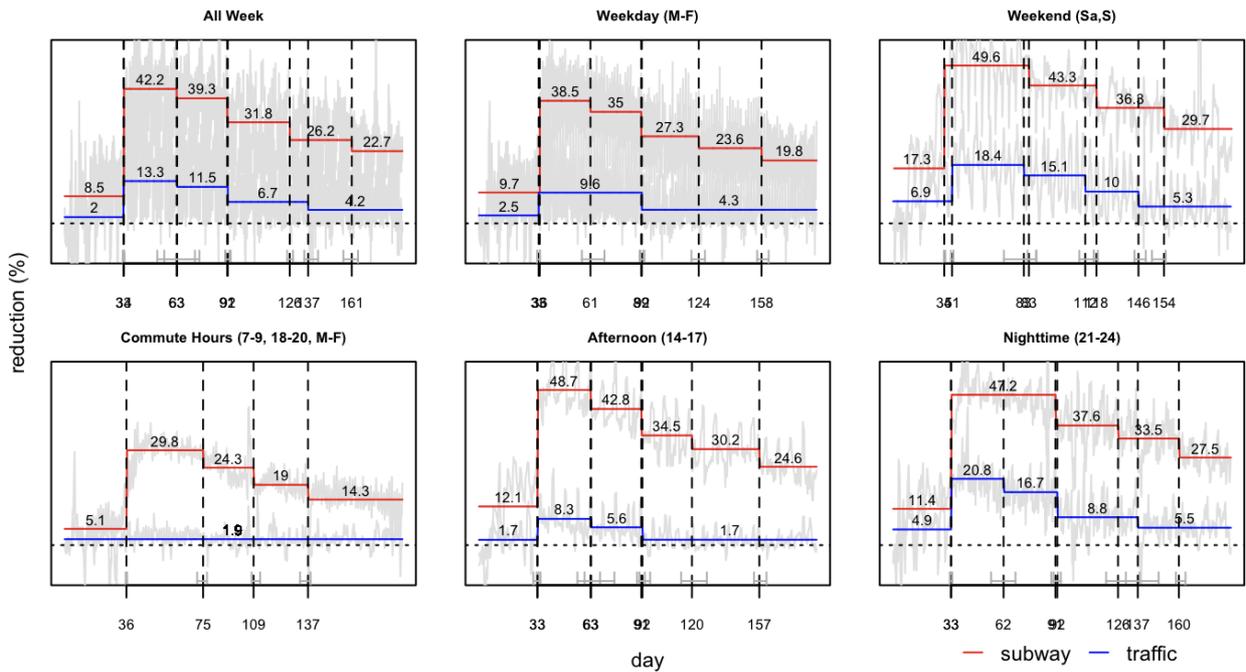

**Figure 3. Time-of-day and day-of week analyses of hourly mobility pattern changes.** All week, weekday, and weekend plots show the day-of-week effect in mobility reductions. Commute hours (7AM-9AM & 6PM-8PM), afternoon (2PM-5PM) and nighttime (9PM-12AM) plots show time-of-day effect.

### 3.4. Mobility, risk perception and universal facemask use

In this section, subway ridership reduction is assessed with the public risk perception, as it is the main trigger of the behavioral adjustment. In Figure 4, bi-weekly public survey outcomes[32] on overall risk of COVID-19 (shown in solid square) vs. risk of personal infection (shown in triangle) are shown. The rate of universal mask use (shown in circle) and subway ridership reduction percentage (shown in blue) are also included. Overall, the personal infection risk perception shares the key characteristics of epidemic phases, as it increased during two escalation phases, remained over 70% during two peaks, and decreased during the de-escalation.

It is notable that the universal facemask wearing had already reached over 60% in escalation-1. Considering the official promotion on universal face covering had started and remained as a recommendation from day 63 (peak-1) to day 217 (peak-2), the public's awareness on the benefit of facemask had been in place much earlier than the government recommendation. As some studies suggested[38], the high compliance level is likely the result of the past infectious outbreak experiences amplified by the current epidemic risk. Since escalation-2, mask use exceeded both the personal infection risk perception and the subway ridership reduction. The mask-wearing experiences in the earlier period seemed to have led the public to embrace the facemask as the longer-term personal safety measure, especially in returning to one's mobility routines.

It is also worth noting that the personal risk perception remained higher than the overall epidemic risk perception in the second wave. The government's social distancing level also stayed at the lowest Level 1[1] during peak-2, since its level is solely determined by the case count trend. When the geospatial momentum is considered, it is evident that Seoul had missed the opportunity to contain the second peak, and the inopportune timing of national economic stimulus to promote in-person shopping and dining since day 106[39]. However, the high level of public's risk perception remained as the main force to prevent rapid resurgence despite the reactive government response.

---

[1] South Korea has adopted a three-level social distancing scheme, where the level of intervention is determined based on weekly average case count and reproduction number within the region. In Seoul Metropolitan area, elevation from Level 1 to Level 2 occurs when the weekly average case count exceeds 40 and the reproduction number is higher than or equal to 1.3 [40]. Elevation from Level 2 to Level 3 occurs when a) two-week average daily toll exceeds 100 and b) daily case count is doubled from the previous day two times within a week, but after c) careful review on the potential impact on the social, economic and medical costs as well as public and expert opinions. More details are included in Supplementary Information.

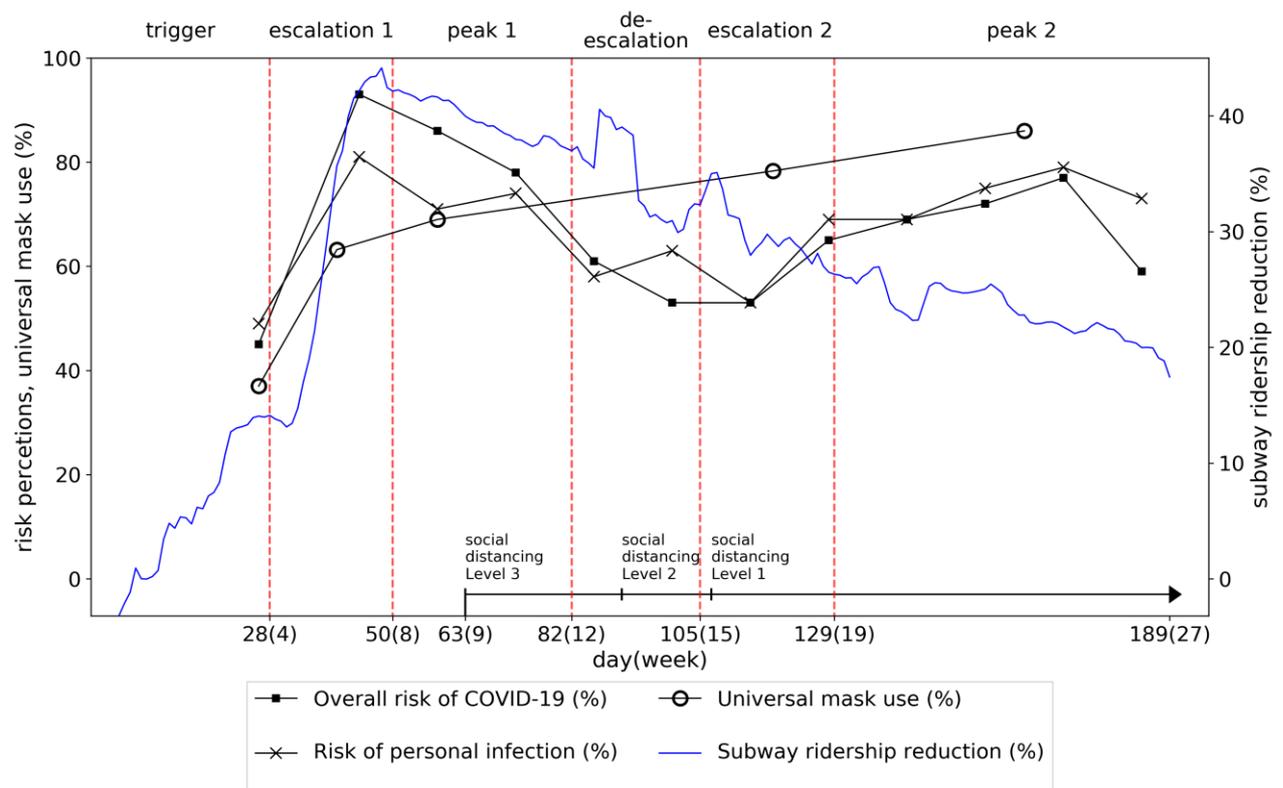

**Figure 4. Survey results on risk perception and mask use by epidemic phases.** Overall risk of COVID-19 is the percentage who responded that the spread of the pandemic within the nation was "very" or "somewhat" serious[32]. Risk of personal infection is the percentage who responded that the likelihood of contracting the virus was more than or equal to 50%[32]. The rate of universal mask use[33-37] is the proportion who responded they have worn masks whenever necessary such as in public space.

## 4. Discussion

In this paper, behavioral changes during COVID-19 outbreak were evaluated in Seoul, South Korea focusing on mobility and facemask use under the marginal mobility restriction. The epidemic phases were defined based on case count and geospatial momentum. Through the phase-specific analyses, we unraveled the true nature of inter-dependency among the epidemic progression, mobility, facemask and public perception. Linear models between mobility reduction and COVID-19 case counts revealed strong linearities during the escalation and de-escalation phases of the first wave, manifesting their interactive nature. The second wave turned out to be not another repetition of the first wave, as mobility continued to recover regardless of fluctuating COVID-19 risk. The public's risk perception showed their vigilance to closely monitor the epidemic progression. As the pandemic persisted for a longer period, the shift in focus from mobility to facemask was evident as the main means of personal safety. The results and interpretations were only visible after establishing the epidemic phases, which demonstrates the significance of multi-faceted data collection and analyses to trace the evolving evidence of the ongoing pandemic.

Although low mobility stringency and gentle nudge on mask-wearing seemed counter-intuitive to fight a pandemic, evidence from South Korea indicated otherwise. Our study demonstrated the self-driven nature of behavioral adaptations with shifted focus from mobility reduction to facemask use, showing their substitutive nature. However, KCDC noted in July briefing that mask-wearing rate fell close to 50% in the indoor facilities such as restaurants and coffee shops, where 2-meter distancing is rarely abided[41]. The recent domestic resurgence from day 207 (week 30) are mostly the clusters originated from close contacts in groups such as in restaurants and churches[42], pushing the contact-tracing near the full capacity. Reflecting on the effectiveness of mobility reduction in the first escalation, our observations from the second peak manifest that facemask cannot be a perfect substitute for mobility restriction. Mobility policy should be the main prevention measure during the escalation and peak phases while the facemasks should serve a key enabler for the fluid mobility recovery during the de-escalation.

COVID-19 is a pandemic we have never experienced in our modern history, but it might not be the last. Learning from the evolving evidence on behavioral adaptations and the resulting public cooperation will have a long-lasting impact to fight this pandemic, and to prepare for the next. However, understanding behavioral changes amid the unprepared pandemics such as COVID-19 is a tall order. As future research, we suggest the following topics for the broader academic communities to explore.

In mobility, our study specifically focused on trip modes in mobility pattern mining, since they are direct observations and have been in service the whole period. From the transportation economics perspective, however, not only the trip mode but also the trip purposes need to be evaluated to fully capture the behavioral aspects, as the mode is a mere means to achieve one's trip purpose. Therefore, expansive data analytics on trip purposes are imperative for adaptive and targeted mobility control. Moreover, analyses on trip purpose may validate the efficacy of various mobility controls, and to identify the high-risk sectors. Incorporating demographic groups such as age, gender and occupations may also broaden our understanding.

In epidemiology, a broader set of metrics might be helpful to monitor and track the progression closer to real-time. In the prolonged battle with COVID-19, the importance of timely response empowered by fast data tracking is no less than searching for the ultimate medical interventions. The changing point detection method and geospatial progression measures effectively captured the progression of the outbreak on multiple ends. Further geometric pattern mining methodology such as topological data analysis[43] may reveal the spatial characteristics even further, which has critical implications in the infectious outbreaks as COVID-19. The necessary condition of data availability and methodological validation must be met, and requires active and collaborative contributions of broader academic disciplines.

In policy, there needs to be a framework to evaluate the efficacy and relevance of key control measures in a systematic and collective manner. Although the testing-contact tracing-isolating has proved successful in South Korea[44], the process of contact tracing had unintended social impacts, especially on the underrepresented populations[45-46]. Moreover, adherence to contact tracing has put growing pressure on the nation's universal healthcare system, and required an enormous sacrifice of the healthcare professionals[47]. Our study outcome strongly suggests that the public is willing to adapt to a reasonable level of mobility restriction, which can alleviate the enduring pressure on the national healthcare infrastructure. Contribution of facemasks and social distancing needs to be considered as a set of policies than of competing interests.

**Data availability**

The data that support the findings of this study except contact tracing information are available at https://doi.org/10.6084/m9.figshare.c.5015093[48].

**Code availability**

Custom code that supports the findings of this study is available from the corresponding author upon request

**Acknowledgements**

This work was supported by the National Research Foundation of Korea (NRF) grant (No. 2020R1A2C2010200). The funders had no role in study design, data collection and analysis, decision to publish or preparation of the manuscript.

**Supplementary Information**

**< List of contents >**

- **Supplementary Note 1.** COVID-19 Epidemiological data description
- **Supplementary Note 2.** Mobility data description
- **Supplementary Note 3.** Survey source description
- **Supplementary Note 4.** Policy responses in South Korea
- **Supplementary Table 1.** Linear fit result
- **Supplementary Table 2.** Changing Point Detection (CPD) Results
- **Supplementary Table 3.** Summary of changes among amendments on *disclosure of itinerary and information of confirmed patients*
- **Supplementary Table 4.** Summary of datasets used in this study
- **Supplementary Table 5**. Summary of government responses for thirteen policy indicators

**Supplementary Note 1.** COVID-19 epidemiological data description

The epidemiological data of the first 189 days consists of two datasets: COVID-19 daily tolls and contact tracing information. COVID-19 daily toll data is collected from KCDC[46], and contact tracing information is sourced from the Kaggle COVID-19 open data[47] for the first 133 days from the Seoul Metropolitan government[48] since day 134. The daily toll includes the number of cases confirmed between 0:00 and 23:59 on the same day. Before day 43, the daily toll referred to cases confirmed between 16:00 of the previous day and 15:59 of the day.

Contact tracing information is the itinerary of a confirmed patient from a few days before the symptom onset to the day of quarantine[49]. If a patient shows no symptoms at the time of quarantine, the itinerary of the patient is collected from a few days before the day of testing to the day of quarantine. The itinerary is recorded with the coordinates of stores, buildings visited, and the means of transportation used each day. In the case of Seoul, the administrative office of each of 25 districts (i.e. *Gu*) collects and publishes the itinerary information of each patient on the administrative office's webpage, and the Seoul Metropolitan government also publishes the data collected from each district office on its official webpage. The final contact tracing data used for the calculation of the geospatial distance of Seoul up to day 189 contains 5,309 contact locations of 1,288 confirmed cases. We have used anonymized contact tracing information provided by KCDC. For data accessibility, please inquire directly to KCDC.

When the outbreak first occurred in South Korea, there was no clear policy on to what extent the information should be disclosed. Before KCDC first distributed the first edition of "*the disclosure of itinerary and information of confirmed patients*" on day 55 (March 14), some privacy-sensitive information was disclosed without strict guidelines. Information released during this period contained travel records of all places such as nightlife venues. The legal guideline for information disclosure was first enacted on day 55 and has undergone two amendments since then. When first published, the guideline defined the purpose of the disclosure and specified the scope and details of information to be released. The guideline stated that information, such as itinerary and close contacts*, should be disclosed to the extent that no individual can be specified; particular places and transportation modes where unknown contacts might have occurred should be informed to the public. Detailed addresses and names of workplaces would not be disclosed unless virus transmission to unspecified employees was expected. This edition stated that each municipal government "may not" disclose the places where the confirmed patient stayed, if all contacts were confirmed.

The second edition of "*the disclosure of itinerary and information of confirmed patients*" was published on April 12 (day 84). This edition stated that contact information should be disclosed only for 14 days after the patient had the last contact. The time range was extended to two days before the date of symptom onset or testing. This edition specified that each government should notify the public about the sterilization status of places visited by patients.

The third edition was announced on June 30 (day 163). This edition emphasized that each patient's information should be deleted 14 days after the patient had the last contact. The guideline for protecting personal information has also been changed so that gender, age, nationality, detailed address (only provide Gu-wise information) of residence, and names of workplaces are not disclosed. Since this edition, information to be released has been in the form of a list of visited places, including region, type, name, and address of the facility, exposed date, and status of sterilization, instead of the itinerary form. In addition, if all contacts in a particular place have been identified, the information of the place visited by a patient "should not" be disclosed to the public. The information of cluster cases is to be announced by KCDC, not by municipalities. Supplementary Table 3 summarizes the changes in "*the disclosure of itinerary and information of confirmed patients*" over the three editions.

The electrical signature check-in system, launched on June 10 (day 143), has indirectly affected the accessibility of contact tracing data. After the implementation of the system, investigators have been able to trace the close contacts of confirmed patients with more completion. As stated in "*the disclosure of itinerary and information of confirmed patients*", each local government "may not" (day 84-162) or "should not" (day 163-ongoing) disclose the place visited by a patient if all the contacts have been identified. As this guideline was changed to "should not reveal such locations" on day 163, there have been more cases that all contacts are identified through an electric signature check-in system, leaving a considerable portion of contact location information unpublished.

The electrical signature check-in system had undergone a pilot period from June 1 (day 134) to June 7 (day 140) and began operation from June 10 (day 143). The period prior to June 30 (day 163) was the pre-implementation period where no legal penalties were imposed. Initially, implementation has been required in eight facilities classified as high-risk, including pubs, nightclubs, colatech (non-alcohol nightclub), karaoke, indoor group exercise facilities, and indoor standing concert hall. On June 12 (day 145), the Seoul metropolitan government required large cram schools and PC cafes to implement the system, and on June 21 (day 154), the KCDC further classified buffet, large cram school, door-to-door sales promotion facility, and logistics distribution center as high-risk facilities.

---

\* *Contacts* indicate people that a COVID-19 confirmed patient may have been in contact with. The scope of contacts is determined considering symptoms of the patient, mask-wearing status, duration of the stay, and exposed situation. All contacts, both symptomatic and asymptomatic, are subject to self-quarantine for 14 days with daily monitoring by public health staff.

**Supplementary Note 2.** Mobility data description

The subway ridership and traffic volume of the year 2019 and 2020 are direct observation counts. Subway ridership data was collected from Seoul Metro[50] and consists of the hourly number of passengers getting on and off at 275 stations. Traffic volume data, consisting of hourly traffic volumes at 270 locations including major road links in Seoul such as downtown areas, city borders, bridges, arterials, and city highways, was collected from Seoul TOPIS[51]. We used traffic volume data of 100 locations with a missing rate of less than 0.25% in the analysis. Missing values were filled in with the median value of the traffic volume of the week before and after. Reductions are measured by matching the closest day-of-week of 2020 to 2019. For instance, Friday, February 7, 2020 was matched with Friday, February 8, 2019. Over the 189 days, there were seven holidays; Lunar New Year (January 24-27), Independence Movement Day (March 1), Buddha's Birthday (April 30), Labor Day (May 1), Children's Day (May 5), the Parliamentary Election (April 15), and Memorial Day (June 6). Holidays were matched with the corresponding 2019 holidays, except that the quadrennial Parliamentary Election day was matched with the closest day-of-week. Both the traffic volume and subway ridership data are provided in figshare[45]. For more details, please refer to our data summary paper[52].

**Supplementary Note 3.** Survey source description

Risk perception rate was obtained from a series of surveys conducted by Gallup Korea[32]. Risk perception rate was measured as the percentage of respondents in Seoul (n=193; aged 18 or over) who rated their level of worry related to coronavirus infection as "very worried" or "somewhat worried". We retrieved the risk perception trend from a total of 16 surveys, which were conducted weekly between February 4 (day 16) and April 8 (day 80) and biweekly between April 21 (day 93) and July 16 (day 179). The survey results have been posted on its official website and are publicly available for research use[32].

The two surveys[33] conducted in early February (day 16) and April (day 74) measured mask adoption rate by investigating the percentage of respondents across the nation (n=2,002; aged 18 or over) who have used facemasks for protection against COVID-19. Survey design and questionnaire are detailed in Jang et al. (2020). Gallup International also surveyed the percentage of respondents across the nation (n=801; aged 18 or over) who have adopted mask-wearing as a precautionary behavior in March (day 61)[34]. In addition, from a series of surveys conducted in late February (day 40) (n=973)[35], May (day 117) (n=1,000)[36], and June (day 162) (n=1,000)[37], we collected the percentage of people across the nation who responded that they have "always" worn facemasks over the past week when outside the home. The late February survey design and results are detailed in Lee & You (2020), and the May, June, and July survey results were retrieved from media releases[36-37].

**Supplementary Note 4.** Policy responses in South Korea

**Mobility restriction.** To back up our statement that the nation has had marginal mobility restriction, we summarize government responses adopted throughout COVID-19 (see Supplementary Table 5). We also quantify the stringency of policy responses, by adopting scoring criteria and method of Oxford COVID-19 Government Response Tracker (OxCGRT)[53]. Specifically, we adopted two indices—government response index and mobility restriction index— to show that mobility restriction has been relatively marginal compared to overall government responses (see figure S1). Government response index, which was first proposed in the working paper of OxCGRT, is based on 13 policy indicators including workplace closure and stay-at-home requirement[9]. Mobility restriction index is calculated based on only 3 indicators that directly constrain mobility: public transport closure, stay-at-home requirement, and internal movement restriction. The types of indicators as well as the scoring method used in this study are solely based on OxCGRT[9]. Scoring codebooks and relevant sources are detailed in Supplementary Table 5.

In summary, the government has not mandated stay-at-home for the public, as of Aug 13. The official advice has been a recommendation on not-leaving-house, even throughout strict social distancing period (day 63-91). Public transport has not been closed during COVID-19. Internal movement restriction was also on recommending to refrain from traveling from/to special disaster zones. Please see Supplementary Table 5 for a detailed description and references.

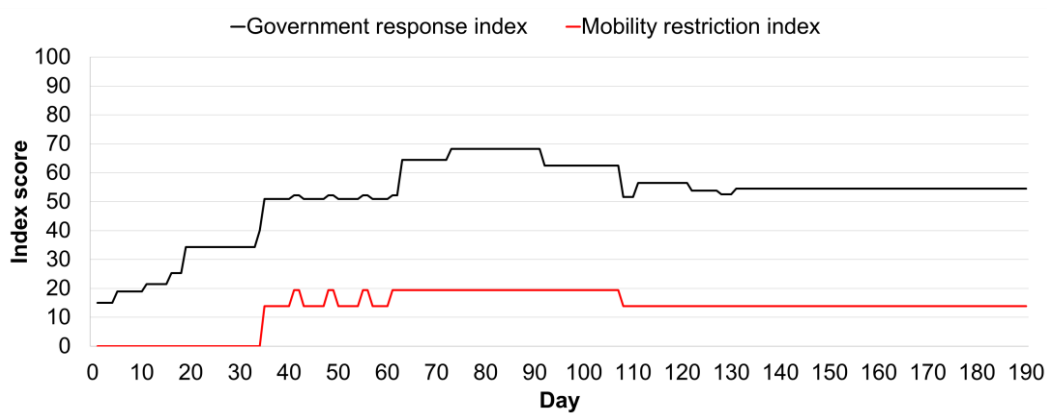

**Figure S1.** Government response index and mobility restriction index, based on OxCGRT

**Facemask policy responses.** KCDC's official advice for mask-wearing before day 63 was that mask-wearing is only recommended for those showing symptoms, those caring for people with symptoms, those visiting medical facilities, those working in environments with a high risk of infection, and those with underlying illness[54]. Only since strict social distancing (day 63), KCDC has recommended the general public to wear masks in crowded indoor spaces[55]. Mask wearing has been also mandatory in public transport in Seoul from day 115 and in the nation from day 128[56].

**Supplementary Table 1**. Linear fit result

| Traffic Volume vs. Number of Confirmed Cases | | | | | | | | | |
|---|---|---|---|---|---|---|---|---|---|
| Phase | | Estimate | Standard error | t | P-value | R2 | Adjusted R2 | F | Significance F |
| Trigger | Intercept | -1.44e-3 | 5.61e-3 | -0.257 | 0.800 | 0.320 | 0.294 | 12.22 | 1.71e-3 |
| | Coeff. | 0.036 | 0.010 | 3.496 | 1.71e-3 | | | | |
| Escalation | Intercept | 0.029 | 3.98e-3 | 7.18 | 5.94e-7 | 0.941 | 0.938 | 317.5 | 9.68e-14 |
| | Coeff. | 0.013 | 7.51e-4 | 17.82 | 9.68e-14 | | | | |
| Peak 1 | Intercept | 0.088 | 7.11e-3 | 12.324 | 2.86e-13 | 0.027 | -5.67e-3 | 0.825 | 0.371 |
| | Coeff. | 4.35e-4 | 4.78e-4 | 0.908 | 0.371 | | | | |
| De-escalation | Intercept | 0.063 | 7.36e-3 | 8.564 | 2.72e-8 | 0.203 | 0.165 | 5.336 | 0.031 |
| | Coeff. | 5.25e-3 | 2.27e-3 | 2.310 | 0.031 | | | | |
| Re-escalation | Intercept | 0.062 | 3.11e-3 | 20.054 | 1.26e-15 | 0.3304 | 0.3 | 10.86 | 3.30e-3 |
| | Coeff. | -1.44e-3 | 4.37e-4 | -3.295 | 3.3e-3 | | | | |
| Peak 2 | Intercept | 0.042 | 3.85e-3 | 10.946 | 9.79e-16 | 0.090 | 0.074 | 5.741 | 0.020 |
| | Coeff. | -6.97e-4 | 2.91e-4 | -2.396 | 0.080 | | | | |
| Subway Ridership vs. Number of Confirmed Cases | | | | | | | | | |
| Phase | | Estimate | Standard error | t | P-value | | Adjusted | F | Significance F |
| Trigger | Intercept | 0.054 | 0.016 | 3.440 | 1.97e-3 | 0.022 | -0.015 | 0.598 | 0.446 |
| | Coeff. | 0.022 | 0.029 | 0.773 | 0.446 | | | | |
| Escalation | Intercept | 0.117 | 0.015 | 7.89 | 1.45e-7 | 0.9157 | 0.9115 | 217.3 | 3.316e-12 |
| | Coeff. | 0.041 | 2.80e-3 | 14.741 | 3.32e-12 | | | | |
| Peak 1 | Intercept | 0.408 | 0.013 | 31.46 | <2e-16 | 0.032 | -6.25e-4 | 0.981 | 0.33 |
| | Coeff. | -8.63e-4 | 8.71e-4 | -0.99 | 0.33 | | | | |
| De-escalation | Intercept | 0.321 | 9.81e-3 | 32.743 | <2e-16 | 0.358 | 0.327 | 11.71 | 2.56e-3 |
| | Coeff. | 0.010 | 3.03e-3 | 3.422 | 2.56e-3 | | | | |
| Re-escalation | Intercept | 0.326 | 7.85e-3 | 41.535 | <2e-16 | 0.450 | 0.425 | 18 | 3.332e-4 |
| | Coeff. | -4.68e-3 | 1.10e-3 | -4.243 | 3.33e-4 | | | | |
| Peak 2 | Intercept | 0.209 | 9.12e-3 | 22.882 | <2e-16 | 0.112 | 0.097 | 7.309 | 8.99e-3 |
| | Coeff. | 1.86e-3 | 6.89e-4 | 2.703 | 8.99e-3 | | | | |

**Supplementary Table 2.** Changing Point Detection (CPD) Results

| Trip mode | Day of Week | Time-of-day | Intervention Estimates: Day | 95% confidence intervals | Fitted Values |
|---|---|---|---|---|---|
| Subway | All week | All Hours (5-24) | 34(3), 63(13), 91(19), 126(19), 161(13) | {34(1), 34(5)}, {56(14), 73(17)}, {91(1), 93(3)}, {125(11), 128(15)}, {156(19), 165(4)} | 8.5, 42.2, 39.3, 31.8, 26.2, 22.7 |
| | | Afternoon Off-peak (14-17) | 33(3), 63(2), 91(3), 120(1), 157(3) | {33(2), 34(2)}, {60(1), 68(3)}, {89(2), 93(3)}, {114(1), 128(2)}, {154(3), 161(3)} | 12.1, 48.7, 42.8, 34.5, 30.2, 24.6 |
| | | Nighttime (21-24) | 33(2), 91(3), 126(3), 160(3) | {33(1), 33(3)}, {89(2), 93(1)}, {120(1), 138(3)}, {159(1), 164(2)} | 11.4, 47.2, 37.6, 33.5, 27.5 |
| | Weekday (M-F) | All Hours (5-24) | 24(5), 43(12), 62(19), 84(19), 108(19) | {24(3), 24(7)}, {40(7), 48(19)}, {62(8), 63(14)}, {82(4), 87(7)}, {107(6), 111(13)} | 9.7, 38.5, 35.0, 27.3, 23.6, 19.8 |
| | | Commute Hours (8-9, 18-20) | 24(3), 53(4), 73(1), 93(4) | {24(2), 24(4)}, {51(3), 55(2)}, {72(2), 75(3)}, {90(4), 94(4)} | 5.1, 29.8, 24.3, 19.0, 14.3 |
| | Weekend (Sa,S) | All Hours (5-24) | 21(2), 33(12), 43(12), 53(12) | {21(1), 21(5)}, {33(8), 34(13)}, {43(5), 43(16)}, {51(16), 53(17)} | 17.3, 49.6, 43.3, 36.3, 29.7 |
| Traffic volume | All week | All Hours (0-24) | 33(21), 63(22), 92(6), 137(6) | {33(13), 34(4)}, {{52(19), 76(10)}, {90(19), 93(23)}, {135(2), 142(19)} | 2.0, 13.3, 11.5, 6.7, 4.2 |
| | | Afternoon Off-peak (14-17) | 33(3), 63(3), 92(2) | {31(2), 35(2)}, {56(1), 76(3)}, {90(3), 96(3)} | 1.7, 8.3, 5.6, 1.7 |
| | | Nighttime (21-24) | 33(2), 62(3), 92(3), 137(3) | {32(3), 34(1)}, {55(3), 69(2)}, {90(3), 94(3)}, {131(1), 149(2)} | 4.9, 20.8, 16.7, 8.8, 5.5 |
| | Weekday (M-F) | All Hours (0-24) | 23(20), 63(6) | {23(6), 24(8)}, {62(17), 64(6)} | 2.5, 9.6, 4.3 |
| | | Commute Hours (8-9, 18-20) | No intervention found | No intervention found | 1.9 |
| | Weekend (Sa,S) | All Hours (0-24) | 22(6), 32(21), 41(23), 49(20) | {22(2), 22(11)}, {29(22), 34(18)}, {41(2), 43(5)}, {49(6), 50(22)} | 6.9, 18.4, 15.1, 10.0, 5.3 |

**Supplementary Table 3.** Summary of changes among amendments on "*disclosure of itinerary and information of confirmed patients*"

|  | **1st edition** (day 55) | **2nd edition** (day 84) | **3rd edition** (day 163) |
|---|---|---|---|
| **Duration of information disclosure** | - | Information is disclosed for 14 days after the patients had the last contact | Information is disclosed for 14 days after the patients had the last contact, and should be deleted after the duration |
| **Temporal scope of information** | From 1 day before symptom onset or SARS-CoV-2 testing to the day of quarantine | From 2 days before symptom onset or SARS-CoV-2 testing to the day of quarantine | From 2 days before symptom onset or SARS-CoV-2 testing to the day of quarantine |
| **Spatial scope of information** | Places where all contacts have been identified may not be disclosed | Places where all contacts have been identified may not be disclosed | Places where all contacts have been identified should not be disclosed |
|  | Detailed address and name of workplace should not be disclosed for privacy *workplace may be disclosed if there is a risk of spreading to unidentified employees | Detailed address and name of workplace* should not be disclosed for privacy *workplace may be disclosed if there is a risk of spreading to unidentified employees | Gender, age, nationality, detailed address and name of workplace should not be disclosed for privacy *workplace may be disclosed if there is a risk of spreading to unidentified employees |
| **Subject of information disclosure for cluster cases** | Municipal governments | Municipal governments | KCDC |

**Supplementary Table 4.** Summary of datasets used in this study

| Category | Region | Period | Frequency | Name | Description | Source |
|---|---|---|---|---|---|---|
| Epidemiology | Nationwide | Jan 20, 2020 – Jul 26, 2020 | Daily | COVID-19 case | Number of confirmed cases; | COVID-19 open data[1]; KCDC[2] |
| | | | | Contact tracing | Contact tracing information of confirmed patients in Seoul | |
| Trip mode | Seoul | Jan 20, 2020 – Jul 26, 2020 | Hourly | Subway ridership | Hourly subway ridership counts extracted from the smart card usage information | Seoul Metro[4] |
| | | | | Traffic volume | Hourly traffic volume at 100 count locations | Seoul TOPIS[5] |

**Supplementary Table 5**. Summary of government responses for thirteen policy indicators

| Category | Day | Description | Authority | Score | Regionality | Source |
|---|---|---|---|---|---|---|
| **School closure** (C1 of OxCGRT) | ~ 34 | Decision not to delay school opening (day 9)<br><br>According to the manual*, it is not recommended to close school because students can spread the epidemic epidemic outside the school. For schools to decide whether to delay opening, they need to consult with infectious disease experts and the Minister of Health and Welfare. (day 11) | Ministry of Education | 0-No measure | - | https://news.kbs.co.kr/news/view.do?ncd=4370426<br><br>http://www.hani.co.kr/arti/society/schooling/926453.html |
| | 35 ~ 121 | Postpone school opening as alert level was raised to 4 (the highest) | Ministry of Education | 3-Require closing all levels | 1-Nationwide | https://www.mk.co.kr/news/society/view/2020/02/187047/ |
| | 122 | Plans to resume school opening for<br>-Grade 12 (day 122)<br>-Grade 1,2,9,11 (day 129)<br>-Grade 3,4,8,10 (day 136)<br>-Grade 5,6,7 (day 141) | Ministry of Education | 2-Require closing (only some levels) | 1-Nationwide | https://imnews.imbc.com/news/2020/society/article/5755259_32633.html |
| | 131 ~ongoing | Limit the number of students attending school to two-thirds in all high schools and to a third in all middle and elementary schools in Seoul Metropolitan area and other cities | Ministry of Education, Provincial Office of Education | 2-Require closing (only some levels) | 0-Specific region | http://news.kbs.co.kr/news/view.do?ncd=4458212 |

| Category | Day | Description | Authority | Score | Regionality | Source |
|---|---|---|---|---|---|---|
| **Workplace closure** (C2 of OxCGRT) | 8~ | As the alert level was raised to 3, the Ministry of Industry distributes standard plans to workplaces and recommends teleconferencing and remote working. | Ministry of Industry | 1-Recommend closing or work-from-home | 1-Nationwide | https://news.mt.co.kr/mtview.php?no=2020013111581196166 |
| | 63-91 | **Strict social distancing (equivalent to level 3 social distancing)**<br>- For public institutions, all but the essential personnel are required to work at home<br>-For the private sector, it is recommended that employees work at home. | Central Disaster and Safety Countermeasures Headquarter (CDSCHQ) | 2-Require closing or work-from-home for some sectors or categories of workers | 1-Nationwide | http://www.mohw.go.kr/react/al/sal0301vw.jsp?PAR_MENU_ID=04&MENU_ID=0403&page=1&CONT_SEQ=353673 |
| | 92~107 | **Social distancing with eased restrictions (equivalent to level 2 social distancing)**<br>- Public institutions are required to adopt work-from-home or staggered shifts for a half of employees<br>- Private companies are recommended to adopt a similar approach as suggested for public institutions | CDSCHQ | 2-Require closing or work-from-home for some sectors or categories of workers | 1-Nationwide | http://ncov.mohw.go.kr/tcmBoardView.do?contSeq=354112 |

|  | 108~ongoing | **Social distancing in daily life (equivalent to level 1 social distancing)**<br>- Public institutions are required to adopt work-from-home or staggered shifts for a third of employees<br>- Private companies are recommended to adopt a similar approach as suggested for public institutions | CDSCHQ | 2-Require closing or work-from-home for some sectors or categories of workers | 1- Nationwide | https://www.mois.go.kr/cmm/fms/FileDown.do?atchFileId=FILE_00093124diwmSHr&fileSn=0 |

| Category | Day | Description | Authority | Score | Regionality | Source |
|---|---|---|---|---|---|---|
| **Cancel public events** (C3 of OxCGRT) | 24 | The need to postpone or cancel the collective event is low, and it is recommended to carry out various events, but with the prevention measures. | Central Disaster Management. Headquarter (CDMHQ) | 0 - No measure | 1- Nationwide | http://ncov.mohw.go.kr/tcmBoardView.do?contSeq=352840 |
|  | 34~ | Recommend not to hold public events and crowd gatherings, including those in small indoor spaces such as religious or outdoor events | CDSCHQ | 1 - Recommend cancelling | 1- Nationwide | https://www.hankyung.com/politics/article/2020022265597 |
|  | 63~91 | **Strict social distancing (equivalent to level 3 social distancing)**<br>An administrative order is issued to prohibit public events and meetings where 10 or more people are involved. Sports events are also suspended. | CDSCHQ | 2 - Require cancelling | 1- Nationwide | http://www.mohw.go.kr/react/al/sal0301vw.jsp?PAR_MENU_ID=04&MENU_ID=0403&page=1&CONT_SEQ=353673 |
|  | 92~107 | **Social distancing with eased restrictions (equivalent to level 2 social distancing)**<br>Mandatory events such as national holidays are set according to the number of people, and it is recommended that any events held in the public and private sectors such as local festivals, exhibitions, briefing sessions, and fairs be postponed or canceled. | CDSCHQ | 1 - Recommend cancelling | 1- Nationwide | http://ncov.mohw.go.kr/tcmBoardView.do?contSeq=354112 |
|  | 108~130 | **Social distancing in daily life (equivalent to level 1 social distancing)**<br>If a large number of people need to gather, free up space to keep a distance of 2m or change the gathering time. | CDSCHQ | No measure | - | https://www.mois.go.kr/cmm/fms/FileDown.do?atchFileId=FILE_00093124diwmSHr&fileSn=0 |
|  | 131~ongoing | **Strict measures re-imposed in the Seoul Metropolitan area**<br>It is recommended that residents of the metropolitan area refrain from going out, meetings, and events that are unnecessary. | CDSCHQ | 1 - Recommend cancelling | 0 – specific region | http://ncov.mohw.go.kr/tcmBoardView.do?contSeq=354773 |

| Category | Day | Description | Authority | Score | Regionality | Source |
|---|---|---|---|---|---|---|
| **Restriction on gathering** (C4 of OxCGRT) | 24 | - When organizing events such as large-scale events, festivals, and exams, the health authorities have prepared recommendations for reference by the host organization | CDMHQ | 0 - No measure | - | http://ncov.mohw.go.kr/tcmBoardView.do?contSeq=352840 |
| | | - The need to postpone or cancel the collective event is low, and it is recommended to carry out various events with sufficient quarantine measures | | | | |
| | 34~ | It is advised to refrain from crowded events in a narrow indoor space, including religious events or outdoor events | CDSCHQ | 1- Recommendation on refraining from gatherings | 1- Nationwide | https://www.hankyung.com/politics/article/2020022265597 |
| | 63~92 | **Strict social distancing (equivalent to level 3 social distancing)** An administrative order is issued to prohibit public events and meetings where 10 or more people are involved. Sports events are also suspended. | CDSCHQ | 4 - Restrictions on gatherings of 10 people or less | 1- Nationwide | http://www.mohw.go.kr/react/al/sal0301vw.jsp?PAR_MENU_ID=04&MENU_ID=0403&page=1&CONT_SEQ=353673 |
| | 93~107 | **Social distancing with eased restrictions (equivalent to level 2 social distancing)** An administrative order is issued to prohibit public events and meetings where 50 or more people are involved in indoor gatherings and 100 or more for outdoors | CDSCHQ | 3 - Restrictions on gatherings between 11-100 people | 1- Nationwide | http://ncov.mohw.go.kr/tcmBoardView.do?contSeq=354112 |
| | 108~ | **Social distancing in daily life (equivalent to level 1 social distancing)** No specific guidelines regarding gathering restrictions except for complying with sufficient quarantine measures | CDSCHQ | 0 - No measure | 1- Nationwide | https://www.mois.go.kr/cmm/fms/FileDown.do?atchFileId=FILE_00093124diwmSHr&fileSn=0 |
| | 111~ongoing | **Strict measures applied to specific places in Seoul Metropolitan area** An administrative order is issued to prohibit gatherings at specific places such as nightlife venues (day 111~ongoing) and karaoke (day 124~175). | Seoul Metropolitan government | 3 - Restrictions on gatherings between 11-100 people | 0-Specified region | http://www.hani.co.kr/arti/area/capital/954825.html |
| | 131~ongoing | **Strict measures re-imposed in the Seoul Metropolitan area** It is recommended that residents of the metropolitan area refrain from going out, meetings, and events that are unnecessary. | CDSCHQ | 1- Recommendation on refraining from gatherings | 0 – specific region | http://ncov.mohw.go.kr/tcmBoardView.do?contSeq=354773 |

| Category | Day | Description | Authority | Score | Region | Source |
|---|---|---|---|---|---|---|
| **Close public transport** (C5 of OxCGRT) | - | No measure | - | 0 - No measure | - | |

| Category | Day | Description | Authority | Score | Region/target | Source |
|---|---|---|---|---|---|---|
| **Stay-at-home requirement** (C6 of OxCGRT) | 35- | It is advised to refrain from moving from/to special disaster zone | KCDC | 1- Recommend not leaving house | 0-specific region | http://www.mohw.go.kr/react/al/sal0301vw.jsp?PAR_MENU_ID=04&MENU_ID=0403&page=1&CONT_SEQ=353089 |
| | 41-42, 48-49, 55-56, 61-62 | It is advised to stay at home during the weekends | KCDC | 1- Recommend not leaving house | 1- Nationwide | https://www.donga.com/news/Society/article/all/20200307/100051616/1  http://www.donga.com/news/List/article/all/20200320/100255858/1 |
| | 63-91 | **Strict social distancing (equivalent to level 3 social distancing)** It is advised to refrain from going out except for daily necessities, medical institution visits, and commuting | CDSCHQ | 1- Recommend not leaving house | 1- Nationwide | http://www.mohw.go.kr/react/al/sal0301vw.jsp?PAR_MENU_ID=04&MENU_ID=0403&page=1&CONT_SEQ=353673 |
| | 92~107 | **Social distancing with eased restrictions (equivalent to level 2 social distancing)** It is advised to refrain from meetings, outings, and events that are not necessary or urgent. | CDSCHQ | 1- Recommend not leaving house | 1- Nationwide | http://ncov.mohw.go.kr/tcmBoardView.do?contSeq=354112 |
| | 108~on going | **Social distancing in daily life (equivalent to level 1 social distancing)** It is advised for the elderly to refrain from going outside except for grocery purchases or visits to medical institutions and pharmacies | CDSCHQ | 1- Recommend not leaving house | 0-targeted group (elderly) | https://www.mois.go.kr/cmm/fms/FileDown.do?atchFileId=FILE_000093124diwmSHr&fileSn=0 |
| | 131~on going | **Strict measures re-imposed in the Seoul Metropolitan area** It is recommended that residents of the metropolitan area refrain from going out, meetings, and events that are unnecessary. | CDSCHQ | 1- Recommend not leaving house | 0 – targeted region | http://ncov.mohw.go.kr/tcmBoardView.do?contSeq=354773 |

| Category | Day | Description | Authority | Score | Region | Source |
|---|---|---|---|---|---|---|

| Category | Day | Description | Authority | Score | Regionality | Source |
|---|---|---|---|---|---|---|
| Restrictions on internal movement (C7 of OxCGRT) | 35 - ongoing | It is advised to refrain from moving from/to special disaster zone | KCDC | **1- Recommend not to travel between regions or cities** | 0- specific region | http://www.mohw.go.kr/react/al/sal0301vw.jsp?PAR_MENU_ID=04&MENU_ID=0403&page=1&CONT_SEQ=353089 |

| Category | Day | Description | Authority | Score | Regionality | Source |
|---|---|---|---|---|---|---|
| International travel control (C8 of OxCGRT) | 1~ongoing | As the alert level of quarantine for infectious diseases was raised from 'interest' to 'attention', quarantine procedures were strengthened for all aircraft passengers. For direct flights from Wuhan, the quarantine officer directly conducts an epidemiological investigation of all passengers | KCDC, Incheon airport | 1-Screening | - | https://www.hankookilbo.com/News/Read/202001211778735559 |
| | 9~ongoing | All immigrants from China were asked to fill out their health status questionnaire and check with a quarantine officer. If symptoms are present, a quarantine investigation is conducted, and suspected patients are immediately quarantined at the discretion of the epidemiological investigator or managed in connection with the local government in charge. | KCDC | 1-Screening | - | https://www.yna.co.kr/view/AKR20200126014151017 |
| | 16~ongoing | All foreigners who have visited or stayed in Hubei Province within 14 days are prohibited from entering. | Government | 3-Ban on arrivals from some regions | - | http://m.korea.kr/news/policyNewsView.do?newsId=148868777 |
| | 74~ongoing | Mandatory 14-day quarantine is required for all international arrivals | Government | 2-Quarantine arrivals from high-risk regions | - | http://ncov.mohw.go.kr/baroView2.do |

| Category | Day | Description | Authority | Score | Geospatial scope | Source |
|---|---|---|---|---|---|---|
| Public information campaigns (H1 of OxCGRT) | ~4 | Updates related to COVID-19 outbreak and suggested prevention measures were uploaded on the official website. Government announcements were communicated via TV and newspaper. | KCDC | 1-public officials urging caution about COVID-19 | | |
| | 5~ | KCDC started publishing daily briefing on its official website, TV and youtube | KCDC, CDSCHQ | 2-coordinated public information campaign (e.g. across traditional and social media) | 1-nationwide | https://www.cdc.go.kr/board/board.es?mid=a20501000000&bid=0015 |

| | 11~ | Emergency SMS sent to the public | Ministry of the Interior and Safety | 2- coordinated public information campaign (e.g. across traditional and social media) | 1- nationwide | https://www.safekorea.go.kr/idsiSFK/neo/sfk/cs/sfc/dis/disasterMsgList.jsp?menuSeq=679 |
|---|---|---|---|---|---|---|
| **Category** | **Day** | **Description** | **Authority** | **Score** | **Geospatial scope** | **Source** |
| Testing policy (H2 of OxCGRT) | ~8 | A person is able to be tested for new coronavirus infectious disease patients if 1) he/she had fever or respiratory symptoms within 14 days after visiting China's Hubei Province (including Wuhan) or 2) he or she contacted with suspected patients and had fever or respiratory symptoms within 14 days | KCDC,MOHW | 1 – Only those who both (a) have symptoms AND (b) meet specific criteria | - | https://www.donga.com/news/Society/article/all/20200206/99569909/2 |
| | 19~ongoing | Even if you do not have a history of visiting China, you will be able to receive a test for 'new coronavirus infection' according to your doctor's prescription. The total cost of inspection is fully supported by the government. If one wants to be tested without a doctor's opinion, he/she is not provided with the cost. | KCDC,MOHW | 3 – open public testing | - | https://www.cdc.go.kr/board/board.es?mid=a20507020000&bid=0019 |
| | 38~ongoing | Drive through testing was first adopted in Seoul Metropolitan area | KCDC, MOHW, Municipalities | 3 – open public testing | - | http://www.hani.co.kr/arti/area/capital/933527.html |

| **Category** | **Day** | **Description** | **Authority** | **Score** | **Geospatial scope** | **Source** |
|---|---|---|---|---|---|---|
| Contact tracing (H3 of OxCGRT) | 1~ | Comprehensive contact tracing conducted by epidemiology investigators | MOHW, KCDC, municipalities | 2 - Comprehensive contact tracing | | https://www.mois.go.kr/cmm/fms/FileDown.do?atchFileId=FILE_00092842VodLEN4&fileSn=0 |
| | 67~ | Comprehensive contact tracing conducted via automated system | MOHW, KCDC, municipalities | 2 - Comprehensive contact tracing | | http://www.korea.kr/common/download.do?tblKey=GMN&fileId=190738345 |

| **Category** | **Day** | **Description** | **Authority** | **Score** | **Geospatial scope** | **Source** |
|---|---|---|---|---|---|---|
| Income support (E1 of OxCGRT) | 73~ | Employment maintenance subsidy (partial government burden on leave/holiday workers) | Ministry of Employment and Labor | 1 - government is replacing less than 50% of lost salary (or if a flat sum, it is | - | https://biz.chosun.com/site/data/html_dir/2020/03/25/2020032501311.html |

| Category | Day | Description | Authority | Score | Geospatial scope | Source |
|---|---|---|---|---|---|---|
| | | | | less than 50% median salary) | | |
| Debt / contract relief for households (E2 of OxCGRT) | 19~ | Expansion of new loans, interest rate discounts, extension of maturity, suspension of premium payment | Financial Services Commission | 1 - Narrow relief, specific to one kind of contract | - | https://biz.chosun.com/site/data/html_dir/2020/02/28/2020022802350.html |

## References (supplementary)